# From time-series transcriptomics to gene regulatory networks: a review on inference methods

**Malvina Marku** [a,*] **and Vera Pancaldi** [a,b,*]

[a]Université de Toulouse, Inserm, CNRS, Université Toulouse III-Paul Sabatier, Centre de Recherches en Cancérologie de Toulouse, Toulouse, France;
[b]Barcelona Supercomputing Center, Barcelona, Spain;
[*]To whom correspondence should be addressed:
malvina.marku@inserm.fr , vera.pancaldi@inserm.fr

## Abstract

Inference of gene regulatory networks has been an active area of research for around 20 years, leading to the development of sophisticated inference algorithms based on a variety of assumptions and approaches. With the always increasing demand for more accurate and powerful models, the inference problem remains of broad scientific interest. The abstract representation of biological systems through gene regulatory networks represents a powerful method to study such systems, encoding different amounts and types of information. In this review, we summarize the different types of inference algorithms specifically based on time-series transcriptomics, giving an overview of the main applications of gene regulatory networks in computational biology. This review is intended to give an updated overview of regulatory networks inference tools to biologists and researchers new to the topic and guide them in selecting the appropriate inference method that best fits their questions, aims and experimental data.

## Introduction

In complex system theory, a system is defined as complex if certain properties, such as nonlinearity, feedback loops, adaptation and non-trivial behavior, emerge from the collective interactions between the system components and the surrounding environment (Artime and De Domenico, 2022). As such, all biological systems, especially molecular systems, are inherently complex, and the global structure and behavior of the system cannot be straightforwardly inferred from the (local) properties of its components. For example, characterizing the connection between the genotype and phenotype, and furthermore pathology, not only requires the identification of the molecules involved in the process and their specific characteristics, but also the ways in which these molecules interact with each other, across spatio-temporal scales. To facilitate the representation and study of such complex systems, their interacting components can be represented as a network, commonly visualized as a graph of *nodes (vertices)* connected by *edges (links)* (Fig. 1). In a molecular network, the nodes represent molecular objects of interest (e.g. genes, mRNAs, transcription factors (TFs)), whereas the edges represent the interactions between them (e.g. protein binding, gene co-expression, TF-target regulation, etc.). Therefore, depending on the types of nodes and edges considered, different molecular networks exist, like protein-protein networks (PPI) (Koh et al., 2012; Pellegrini et al., 2004; Schwikowski et al., 2000; Vazquez et al., 2003), gene regulatory networks (hereafter, GRN) (Emmert-Streib et al., 2014; Vijesh et al., 2013), signal transduction networks (Hu et al., 2021; Kolch et al., 2015), etc.. Strictly speaking, we refer to GRNs as networks of any types of regulatory interactions between regulatory and target molecular entities (miRNAs-targets, RBP-targets, kinases/ phosphatases-substrates). Here we will focus mostly on GRNs that describe interactions between TFs and their target genes.

In studying GRNs two main approaches of extracting information exist: (i) *static network analysis*, and (ii) *dynamical modeling*, each of which offers different amounts and types of information regarding the network organization, topology and behavior (He et al., 2015; Jungck and Viswanathan, 2015). Inferring and modeling these regulatory networks, however, is a challenging reverse engineering process and requires the combination of both a thorough biological understanding of the system, and accurate and advanced computational inference methods (Angelin-Bonnet et al., 2019). Moreover, the advanced technological improvements in measuring gene expression, in cellular populations or even single cells, and the increasing interest in clinical applications of genomics, confer importance and relevance to data-driven GRN inference methods. These approaches can ultimately provide considerable insights on gene regulation mechanisms, drugs' mode of action, pathway perturbation, etc. than the original data alone.

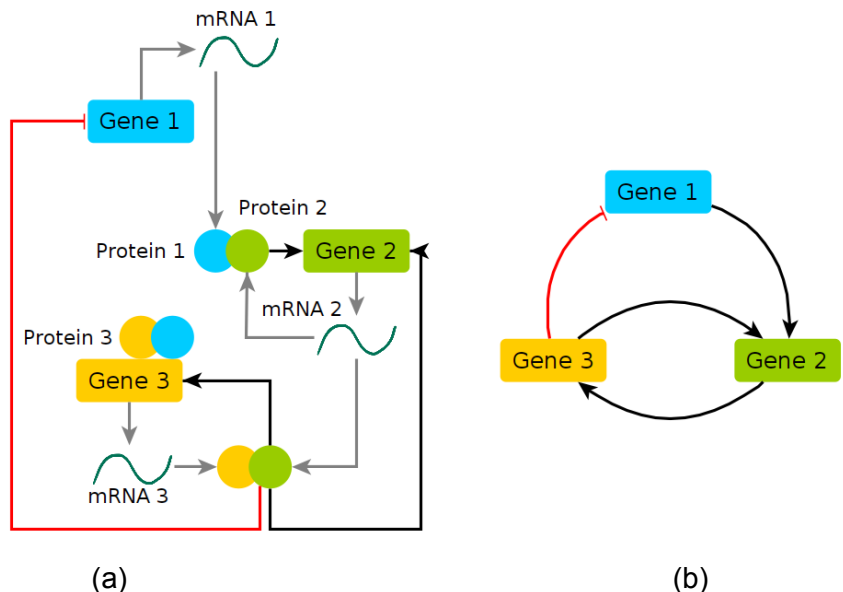

(a)            (b)

**Fig. 1:** (a) A toy model of gene regulation of three genes involved in a transcriptional regulatory network, showing genes transcribed into mRNAs and translated into proteins that regulate another one of the genes. (b) Compressed representation of the interactions on the left as a GRN in which only genes are shown with their regulatory interactions.

With the ever continuous advances in biotechnologies, inference methods have also evolved from using bulk gene expression data (Barabási et al., 2011; Marbach et al., 2012; Silverman et al., 2020; Sonawane et al., 2019) to single-cell transcriptomics (Blencowe et al., 2019; Fiers et al., 2018; Kang et al., 2021, 2021; Matsumoto et al., 2017; Nguyen et al., 2021; Raharinirina et al., 2021a). Additionally, some methods were adapted to infer the regulatory networks from time-series and/or pseudo-temporal single-cell transcriptomics, where more accurate knowledge on gene-gene interaction can be inferred (Aubin-Frankowski and Vert, 2020a; Huynh-Thu and Geurts, 2018; Matsumoto et al., 2017).

Consequently, high performing algorithms have been developed over the years, aiming to increase their accuracy, robustness and applicability, for example incorporating additional computational analyses, such as filtering for the presence of TF motifs in promoters of their inferred targets (Aibar et al., 2017; Alvarez et al., 2016). It is important to note, however, that GRN inference is not the final aim. Our goal in performing such calculations is to produce relevant insight on the biological processes underlying them, aiming to uncover new functionally important molecular regulatory interactions or propose new drug targets, etc (Emmert-Streib et al., 2014). Additionally, network inference offers the possibility of applying dynamical models to GRNs, in order to increase the model power to understand and predict the system temporal behavior (Angelin-Bonnet et al., 2019).

Inference of data-driven accurate and powerful GRNs remains an open and evolving computational challenge, and new inference methods are being published continuously. In this review, we give a description of GRNs and describe the algorithms behind a selection of the available methods on network inference from experimental data. While several reviews on GRN inference based on different data sources have been published (Delgado and Gómez-Vela, 2019; Huynh-Thu and Sanguinetti, 2018; Mercatelli et al., 2020), here we will focus specifically on time- and pseudo-time series transcriptomics based inference algorithms. We conclude by giving an introduction to dynamical modeling of GRNs, as a promising tool to use GRNs in the generation of new biological hypotheses.

## 1. Gene maps: network representations of gene regulation

Although a single definition of GRN does not exist, we define the GRNs as topological maps representing relationships and interactions between biological entities. The interactions between proteins, transcription factors (TFs) and genes can be represented as directed graphs of *nodes* and *edges*. Notably, the directionality of the edges (defining the source and the target in the interaction) is not compulsory in all biological networks, but it represents an important feature in the case of the regulatory networks, defining the direction of information flow, as it is the case in GRNs or metabolic networks. In these networks, additional information is added by indicating specific types of interactions, represented by singed edges (Fig 1, b). In the simplest case, the interactions are categorized as *activations and inhibitions*, represented as *positive and negative edges* accordingly. GRNs are composed of *regulatory nodes* (*source/cause nodes*) and *regulated nodes* (*target/effect nodes*), generally mapped as TF-target gene network, with the incoming connectivity (in-degree) estimated using gene-centered approaches, and the outgoing connectivity (out-degree) using TF-centered approaches.

The structure of the network enables the calculation of various quantities that capture different features of the network topology that can reveal important information on the underlying biology of the system. Centrality measures, such as eigenvector centrality, Page/Chei Rank, Burt's constraint, or alpha centrality have been shown to highlight key nodes in a network based on network structure (Ashtiani et al., 2019, Cohen and Havlin, 2010; Newman, 2010; Newman et al., 2006). Particularly of interest, in combination with mathematical methods such as Signal Flow Analysis and Feedback Vertex Set Control (Lee and Cho, 2018), the network topology can give valuable information on the determinant nodes that define the system temporal behavior (Marazzi et al., 2022).

## 2. Inference methods

By definition, the process of inferring the network structure of the system, based on experimental data is a reverse engineering process, usually referred to as GRN inference. In general, based on input data used, GRN inference methods can be categorized in two types: (i) *steady state gene expression*, and (ii) *time-series gene expression* inference methods. In the first category, in principle, the GRN inference is obtained by considering perturbations of the system or in different instances and estimating the gene expression after it reaches an equilibrium. In the second category, the input data consists in gene expression measured at several time points after a perturbation thus a temporal evolution of transcriptomics can be obtained. Consequently, time series inference methods can be more informative than static data in a wide range of situations to infer gene functionalities, interactions and causal relationships, as well as to establish potential clinical implications of gene expression dynamics determined by these relationships (Bar-Joseph et al., 2012). Both methods, however, have various limitations mostly stemming from technical issues that arise from the experimental protocols: sampling time points, cost, cell synchronization, sparsity of gene expression data, etc. Therefore, several computational methods that combine both steady-state and time-series approaches have been developed, usually based on advanced machine learning algorithms. Additionally, new technology providing expression at the single-cell level have led to the development of inference methods that are specifically adapted to single-cell transcriptomics (Chen and Mar, 2018).

Inferring the functional relationships between genes requires the estimation of gene functional dependencies, which can be broadly achieved by two different reverse engineering approaches for GRN inference:

1. *Model-free methods:* In this approach, gene dependencies are inferred from using several statistical and machine learning methods, such as mutual information (Faith et al., 2007; Margolin et al., 2006; Meyer et al., 2007), random forest (Huynh-Thu et al., 2010; Kimura et al., 2020; Park et al., 2018; Petralia et al., 2015), deconvolution (Chen et al., 2014; Feizi et al., 2013), or epigenetic (Sonawane et al., 2021).
2. *Model-based methods:* In this approach, a quantitative dynamical model (for example, ODEs (Aalto et al., 2020; Huynh-Thu and Geurts, 2018; Iglesias-Martinez et al., 2016), regression methods (Michailidis and d'Alché-Buc, 2013), or bayesian reasoning (Young et al., 2014)) is defined to model the dynamical properties of the system, while the regulatory network is inferred from optimizing the model parameters based on the time-series data. In this way model-based GRN methods highlight some dynamical features of the system, increasing the model interpretability, especially from a biological point of view.

Depending on the scope and biological question for which the GRN is required, model-free or model-based inference methods can be applied; however, it is important to note that model-based methods are considerably more computationally intensive and might be limited by the underlying non-realistic linear models of gene expression dynamics. On the other hand, the *Dialogue on Reverse Engineering Assessment and Methods* (DREAM) project, an initiative to benchmark multiple inference methods, indicated that no single method can perform the best in every possible dataset and across different settings. Instead, a high confidence consensus network inferred from different methods is the most accurate and also provides an estimation of dataset and method robustness and performance - a process that has been usually referred as *Wisdom of Crowds* (Marbach et al., 2012). For this reason, instead of using a single inference method, other researchers have been developing computational tools combining several GRN inference methods (Manica et al., 2021), providing a ranking of the methods according to their performance.

The increasing demand for improving the inference accuracy in real datasets has warranted the adaptation of some steady-state inference methods to consider time-series data. This led to the development of new promising inference methods, whose applications can highlight novel results in various molecular systems. In the next sections we will elaborate on this category of inference tools specifically and give an overview of different algorithms and their performance.

### 2.1. Inference of GRNs from time-series transcriptomics

Inference of gene networks from time-series data provides a more complete picture of the system than steady state data, as the biological system is intrinsically complex but biological function relies on coordinated sets of genes whose expression evolves over time. In general, inference methods based on time-series transcriptomics use a dataset containing a list of genes with their expression measured at several time points. Let's define the dataset as $D_{TS}$, a matrix with dimensions $N \times TP$, where $N$ is the number of genes and $TP$ is the number of time points at which their expression is measured:

$$D_{TS} = \{X(t_1), X(t_2), ..., X(t_p)\} \tag{1}$$

where $X(t_p)$, $p = 1, 2,..., TP$ is a vector of $N$ genes with their expression at time $t_p$. The main goal of inference methods is to assign a weight $w_{j,i} \geq 0, i, j = 1, 2,..., N$ to any putative interaction between gene $i$ (target) and $j$ (source), representing a regulatory interaction in the biological system. To this purpose, different inference methods use various regression tools to model the expression of a gene as a function of its regulators. It is important to note that, usually, the time-series inference methods combine the time-series with steady-state datasets, in order to increase the accuracy and predictive power of the inferred network. Independent of the method chosen, the goal is to reconstruct the GRN that would produce the observed profile of expression across time, in the form of a directed graph, in which each edge is associated with its characteristic weight (Fig. 2).

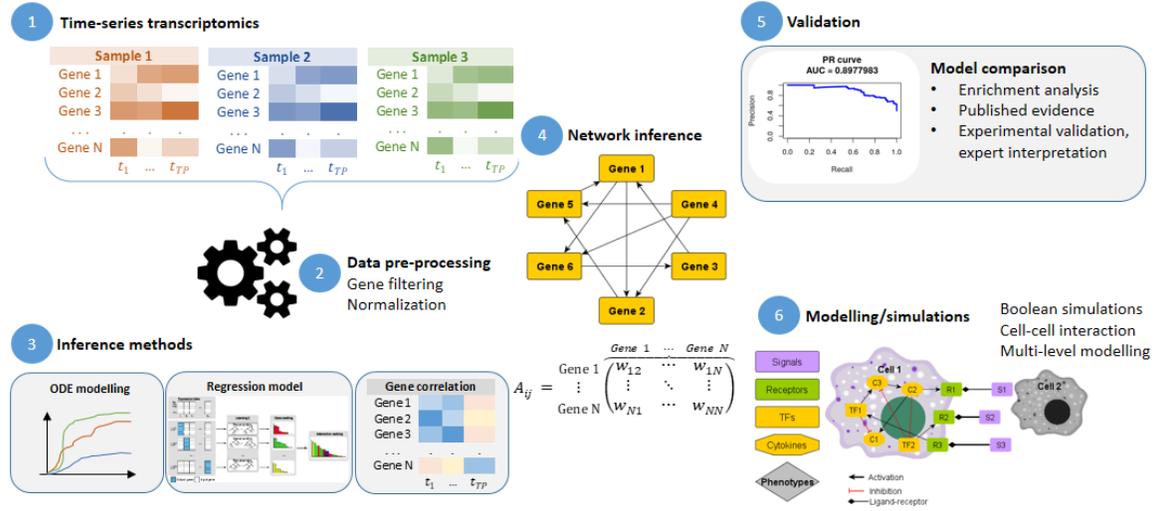

**Fig. 2. From time-series transcriptomics to GRN inference and model validation.** The process usually starts with processing the time-series datasets and identifying the time-points relevant to describe the biological process (1). As a second step, data formatting, normalization and gene filtering and/or binarization might be necessary (2), depending on the inference method to be used (3). The inferred network consists of a weighted directed GRN (4). Model validation (5) is then necessary for quantifying the method performance in inferring the GRN when compared to gold-standard datasets, evidence from literature or prior biological knowledge. The inferred network can further be used for the application of dynamical models in cellular or multilevel modeling (6).

### 2.2. Inference algorithms

Based on the inference model they use, inference methods can be grouped into 7 categories, namely (i) mutual information (MI), (ii) dynamical Bayesian, (iii) Granger causality, (iv) Boolean, (v) ordinary differential equation (ODE), (vi) graphical Gaussian, and (vii) regression. It is important to note, however, that many methods apply a combination of different models and novel approaches. In the following sections, we will introduce the basic concept of each inference algorithm, while some of the available tools for using each of the methods are summarized in Table 1.

*Mutual information based models:* In these methods, the putative interaction between two genes is inferred by estimating the mutual information (MI) between the expression $X_i$ of gene $i$ at time point $t_p$ and the expressions $X_j$ of gene $j$ at the $m^{th}$ previous time points:

$$I_m(g_i, g_j) = \sum_{p=m+1}^{TP} P(X_i(t_p), X_j(t_p - m)) \, log \, \frac{P(X_i(t_p), X_j(t_p-m))}{P(X_i(t_p)) \, P(X_j(t_p-m))} \qquad (2)$$

where $P(X_i(t_p), X_j(t_p − m))$ is the joint probability of observing gene $i$ at time point $t_p$ with expression $X_i(t_p)$ and the expression $X_j(t_p − m)$ of gene $j$ at time point $t_p − m$, whereas $P(X_i(t_p))$ and $P(X_j(t_p − m))$ are the marginal probabilities of observing gene $i$ at time point $t_p$ with expression $X_i(t_p)$ and the expression $X_j(t_p − m)$ of gene $j$ at time point $t_p − m$ independently. Estimating $I_m(g_i, g_j)$ for each pair of genes and each time point, an interaction $g_j \to g_i$ is being considered if $I_m(g_i, g_j)$ exceeds a defined threshold. Among the methods using MI algorithm for GRN inference we highlight TimeDelay-ARACNE (Zoppoli et al., 2010), MRNET (Liu et al., 2017, Meyer et al., 2007) and CLR (Faith et al., 2007). One of the main advantages of these methods lies in their simplicity and speed of computation. On the other hand, their main limitation is that they do not give information on the nature of interactions (activation or inhibition) for the inferred interactions.

***Dynamical Bayesian network models:*** Dynamical Bayesian network model (DBN) GRN inference methods from time-series datasets (Friedman and Koller, 2003; Murphy and Mian, 1999) come as an extension of simple Bayesian network models (BN) applied to steady-state datasets, in order to overcome the limitation of prohibiting the existence of feedback loops in the network. Given that a GRN is defined as a set of vertices and edges $G = (V, E)$, these methods search for any relationship between the expression of a target gene (defined as a random variable) and expression of its regulators (defined as parent genes) by calculating a joint probability distribution. In a time-series experiment, consisting of $N$ genes and $TP$ timepoints, the joint probability distribution of each gene is defined as:

$$P(X1(t_1), X_1(t_2),..., X_1(t_{TP}),..., X_N(t_1), X_N(t_2),..., X_N(t_{TP})) = P(X(t_1)) \times P(X(t_2)|X(t_1)) \times ... \times P(X(t_{TP})|P(X(t_{TP} − 1))$$
(3)

where $X(t_p), p = 1, 2,..., TP$ is a vector representing the gene expression values at time point $t_p$.

The network $G$ is then inferred by identifying the network structure with the highest posterior probability of each edge $E$ from the data, assuming a linear dependency between previous and current expression of genes. The network $G$ is Markovian, meaning that each gene is dependent only on the regulation by its regulators immediately upstream (parents). An edge is then included in the network if the marginal posterior probability of an observation (left hand side of Eq. 3 when written for each gene) exceeds a given threshold.

Despite having a generally good performance when benchmarked against other time-series GRN inference methods (DREAM5 challenge (Marbach et al., 2012)), DBNs are generally computationally expensive and usually limited to small networks. To overcome this limit, several methods like CAS (Xing et al., 2017) or scanBMA (Young et al., 2014, implemented as a function in networkBMA R package (Fraley et al., 2014)), were developed, by combining DBN with machine learning methods or prior knowledge, to guide the search for each gene's regulators.

***Granger causality:*** in these methods, two assumptions in defining the expression $X_i$ of a gene $i$ are made: (1) the expression $X_i$ of gene $i$ at time point $t_p$ is a function of its expression at the $m$ previous time points only, (2) the expression $X_i$ of gene $i$ at time point $t_p$ is a function of both its expression at the $m$ previous time points and the expression $X_j$ of genes $j$ at the $m$ previous time points, thus indicating the effect of gene $j$ on gene $i$. If the second assumption is significantly more successful than the first assumption, it is said that gene $j$ Granger causes gene $i$.

$$X_i(t_p) = \alpha_0 + \alpha_1 X_i(t_p - 1) + \ldots + \alpha_m X_i(t_p - m) + \eta(t_p) \quad (4a)$$

$$X_i(t_p) = \alpha_0 + \alpha_1 X_i(t_p - 1) + \ldots + \alpha_m X_i(t_p - m) + \beta_0 + \beta_1 X_j(t_p - 1) + \ldots + \beta_m X_j(t_p - m) + \eta(t_p)$$
$$i, j = 1, 2, \ldots, N \quad p = 1, 2, \ldots, TP \quad (4b)$$

where α, β are coefficients and η($t_p$) is the residual noise at time $t_p$. Generalizing for $N$ genes, gene $j$ ($j = 1, 2, \ldots, N, j \neq i$) is said to be casual for gene $i$ if considering the expression of gene $j$ in the previous time points significantly improves the prediction of the expression of gene $i$ at the current time point. Notably, when referring to the '*previous time points*', several previous time points ($t_p - 1$, $t_p - 2$, etc) can be considered. Considering $N$ genes in the dataset, the *vector autoregressive (VOA) model* is used to estimate the Granger causality over the entire list of genes, where a linear dependency between genes is assumed (Tam et al., 2013).
Several methods using Granger causality for GRN network inference have been developed, such as BETS (Lu et al., 2021), SWING (Finkle et al., 2018), CGC-2SPR (Yao et al., 2015), among others. One of the notable advantages of this approach is the computational efficiency and speed, compared to other inference methods, with a similar performance. However, considering their definitions and associated algorithms, Granger causality inference methods can be used only on time-series datasets, which are often limited by the sparsity and non-uniformity of time-point spacing. For this reason, the method was extended to consider a flexible time lag between consecutive time-points (Finkle et al., 2018).

***Boolean models:*** in these methods, a GRN is represented as a graph $G = (V, E)$ and $< i, j, s >$ indices, in which $i$ represents the target and $j$ represents the source gene. Each edge is characterized by a sign $s \in \{+, -\}$ indicating the type of interaction between the regulator and the source node. Generally, a positive sign indicates a positive (activating, promoting) relationship between a regulator and its target gene (i.e the regulator gene $j$ contributes in increasing the expression of target gene $i$ through specific biological processes), whereas a negative sign indicates a negative (inhibiting, repressing) relationship (i.e the more expressed the regulator gene $j$ is, the less expressed the target gene $i$ will be). Notably, the expression of each gene is given in Boolean (binary) values $\{0, 1\}$, thus providing only a qualitative description of gene regulation and expression. Given a time-series dataset, the expression $X_i$ of gene $i$ at time $t_p$ is given by a Boolean function of its regulators:

$$X_i(t_p) = F_{B_i}(X_{i,1}(t_p - 1), X_{i,2}(t_p - 1),..., X_{i,k}(t_p - 1)) \quad (5)$$

where $X_{i,k}$, $k = 1, 2,..., N - 1$, $k \neq i$ are the regulators of gene $i$, and $F_B$ is a Boolean function describing the regulation of gene $i$ using the Boolean operators AND, OR and NOT. In this way, the inference of any causal relationship is done by a discrete dynamical model of gene expression (Kauffman and Kauffman, 1993; Shmulevich et al., 2002a).

Compared to the other inference methods, BN inference methods offer the advantage of being parameter free in the modeling of gene regulation and relatively easy to apply. On the other hand, they pose several difficulties and limitations arising from the discretization/binarization process, the finite space explored for inferring the regulatory processes (limited number of possible BNs), and the qualitative description of expression (Pušnik et al., 2022). Nevertheless, several inference methods based on BN have been developed, such as REVEAL (Liang et al., 1998) and Best-Fit (Lähdesmäki et al., 2003; Shmulevich et al., 2002b), both available as extensions in the BoolNet R package (Müssel et al., 2010), and ATEN (Shi et al., 2019).

***Ordinary differential equation (ODE) models:*** in this formalism, the variation of expression $X_i$ of gene $i$ at time $t_p$ is given by a nonlinear function of its regulators (including self regulation) at the same time steps:

$$\frac{dX_i}{dt} = \sum_{k=1}^{N-1} f(X_{i,k}) + \eta_i \quad (6)$$

where the functions $f(X_{i,k})$ are usually assumed to be polynomial functions (Kim et al., 2007) and $\eta_i$ represents the non-deterministic term (noise) in the regulatory function of $X_i$. The regulatory network is inferred solving a system of ODEs (one equation for each gene). From its definition, solving the ODE system (Eq. 6) is computationally and conceptually challenging, even in a linear space, as these systems are defined by a large number of interacting coefficients, for which the prior information is limited or missing. Consequently, ODE methods are limited to inference of small regulatory networks and are usually combined with other inference methods in order to reduce the computational complexity (Margolin et al., 2006; Meyer et al., 2007). In principle, a causal relationship between two genes is considered if the interaction is described by a relatively high interacting coefficient. Despite the difficulty of applying them to real datasets, several methods using ODE models have been developed, including Inferelator (Bonneau et al., 2006) and TSNI (Bansal et al., 2006), but they often display a relatively lower performance than alternative approaches, as reported in (Lu et al., 2021).

***Regression models:*** generally, in these methods a non-linear regression or ODE formulation is used to model the expression of a gene at a certain time point/condition as a nonparametric function of the expression of other genes at the same time point/condition:

$$X_i = f_i(X_j(t_p)) + \eta(t_p), \quad (7)$$

where $\eta(t_p)$ is a random noise term for the $t_p^{th}$ time point. The inference of $f_i$ functions is then done by implementing a feature selection algorithm, like LASSO, least angle regression (LARS), or random forest algorithm to "learn" these functions from an ensemble of regression trees (Breiman, 2001). To identify the candidate genes for the regulation of $N$ genes, this approach starts by splitting the problem into $N$ distinct sub-problems, thus assuming that each gene can be a target gene and a potential regulator. Causal interactions are quantified by estimating the confidence levels of each $j \rightarrow i$ interaction, which is represented by the weights $w_{j,i}$. Importantly, in order to increase the accuracy of the inferred network, some regression methods have been adapted to use both time-series and steady state datasets. Some inference methods based on decision-tree algorithms can be highlighted to have particularly good performance: dynGENIE3 (Huynh-Thu and Geurts, 2018) and Jump3 (Huynh-Thu and Sanguinetti, 2015), both being extensions of the steady-state method GENIE3 (Huynh-Thu et al., 2010) for time-series datasets.

***Gaussian process (GP) models:*** in these methods, a GP regression is used to model the relationship between the current expression of the genes and their previous expression levels, modeling the functional relationships between targets and their regulators as Gaussian processes:

$$X_i(t_p) = f(X_j(t_p)) + \eta(t_p) \tag{8}$$

A causal relationship (edge) between two genes is considered based on the sum of its posterior probabilities of the existence of this relationship (edge) over the time points. In this way, Gaussian process models are generally considered as nonlinear DBN. Consequently, the derived inference methods, such as the algorithm presented by Aijo and Lähdesmäki (Aijö and Lähdesmäki, 2009) and BINGO (Aalto et al., 2020) combine both DBN and GP.

### 2.3. Inference from time-series of single-cell transcriptomics

For some biological processes, like cell differentiation, phenotypic reprogramming, etc, a higher resolution of the temporal dynamics of gene expression is necessary to identify major phenotypic transitions in complex tissues while characterizing the phenotypic spectrum of individual cells. In this regard, single-cell RNA-sequencing technology enables deeper investigation of the molecular interactions and identification of novel molecular mechanisms that orchestrate biological processes at the single cell level. Computationally, this technological revolution has led to the development of several algorithms to analyze single-cell RNA-seq, and - as a part of it - inference of GRNs (Fig. 3). Intuitively, a single-cell resolution of genes' dynamics would lead to an increased accuracy in inferring the functional interactions between genes that define the biological process. However, dealing with limitations in some of the most widely available single-cell technologies, the heterogeneity and sparsity of single-cell data lead to limitations and challenges for GRN inference methods and put reliability in question.

An important point in using single-cell RNA-seq data for GRN inference is - usually - the lack of time-resolved expression measurements. Instead, many inference methods exploit the multiplicity of RNAseq profiles at one single time point across cells as a proxy for temporal evolution of the phenotype, as is the case for trajectory inference based on pseudo-time ordering of the cells ([Aibar et al., 2017](#)), assuming ergodicity of the phenotypes.

**Table 1:** Tools for GRN inference from time-series transcriptomics categorized by their inferring algorithm. The characteristics of the inferred network are indicated as follows: ⊘ undirected, ▷ directed and unsigned, ▶ directed and signed.

| Tool | | Method | Language | Reference |
|---|---|---|---|---|
| Time-delayed ARACNE | ▶ | Mutual Information | R | ([Zoppoli et al., 2010](#)) |
| MRNET | ⊘ | Mutual Information | `minet` R package | ([Liu et al., 2017](#)) |
| CLR | ▷ | Mutual Information | `minet` R package | ([Faith et al., 2007](#)) |
| Jump3 | ▷ | Regression | Matlab/ R | ([Huynh-Thu and Sanguinetti, 2015](#)) |
| dynGENIE3 | ▷ | Regression (+ ODE) | R | ([Huynh-Thu and Geurts, 2018](#)) |
| SWING-RF | ▷ | Granger causality + Regression | Python | ([Finkle et al., 2018](#)) |
| BETS | ▷ | Granger causality | Python | ([Lu et al., 2021](#)) |
| CGC-2SPR | ▷ | Granger causality | R/Matlab | ([Yao et al., 2015](#)) |
| CAS | ▷ | Dynamical Bayesian Model | No info | ([Xing et al., 2017](#)) |
| scanBMA | ▷ | Dynamical Bayesian Model | R | ([Young et al., 2014](#)) |
| Package `GeneNet` | ▷ | Dynamical Bayesian Model | R | ([Kolpakov et al., 1998](#)) |
| REVEAL | ▷ | Boolean Network + Mutual Information | C | ([Liang et al., 1998](#)) |
| BoolNet | ▷ | Boolean Network | R | ([Müssel et al., 2010](#)) |
| GABNI | ▶ | Boolean Network + MI + Genetic Algorithm | | ([Barman and Kwon, 2018](#)) |
| ATEN | ▶ | Boolean Network + Regression | R | ([Shi et al., 2019](#)) |
| TSNI | ▶ | ODE | Matlab | ([Bansal et al., 2006](#)) |
| Inferelator | ▶ | ODE | R | ([Bonneau et al., 2006](#)) |

| BINGO | ▷ | Gaussian process | Matlab | (Aalto et al., 2020) |

In this process, the pseudo-temporal trajectory is generated by linearly ordering the single-cell profiles from a specific time point based on their transcriptional similarity, thus enabling the identification of gene patterns along the developmental trajectory of continuously ordered cells (Zeng et al., 2017). Accordingly, a subset of the inference methods require specific information about the pseudotemporal ordering of the cells (SCODE (Matsumoto et al., 2017), SINCERITIES (Papili Gao et al., 2018), SINGE (Deshpande et al., 2022), LEAP (Specht and Li, 2016), SCRIBE (Qiu et al., 2020), etc.), having a significant difference in performance when such information is not available. Other methods, like GENIE3 (Huynh-Thu et al., 2010), GRNBoost2 (Moerman et al., 2019), or PPCOR (Kim, 2015) do not require a temporal ordering of the cells as input and have relatively good performance when tested on some published curated models (Pratapa et al., 2020). However, the incomplete equivalence between bulk gene expression time courses and pseudotime time-series implies that these two types of inference cannot be always performed by the same tools.

Several benchmarking papers on the performance of single-cell RNA-seq inference methods have been published, facilitating the benchmarking of different inference methods. We refer the reader to (Blencowe et al., 2019; Fiers et al., 2018; Nguyen et al., 2021) for an extensive review and comparison of single-cell RNA-seq inference methods, (Bellot et al., 2015; Kang et al., 2021; Pratapa et al., 2020) for some benchmarking libraries and to (Huynh-Thu and Sanguinetti, 2015) for an algorithmic review. In principle, these methods follow a similar classification as in bulk transcriptomics, described in Section 2.2, as many of them have been adapted for usage in single-cell from previous existing methods for bulk data. For this reason, here we simply give a list of some of these methods (Table 2), noting that there are many more and the list is increasing rapidly.

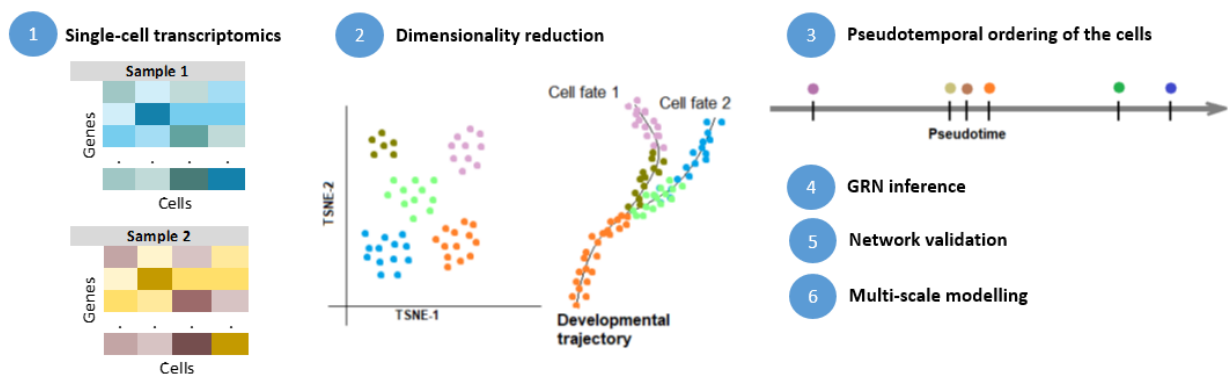

**Fig. 3: GRN inference from single-cell (pseudo)time- series.** The inference workflow follows a similar path as in bulk time-series transcriptomics, except from additional steps of dimensionality reduction and trajectory inference (2), and pseudotime ordering of the cells (3) when time-resolved experimental measurements are not available. Steps (4) - (6) follow the same logic as in Fig. 2.

For a user, the choice between all the different inference methods will depend on both their overall performance, the type and amount of information they require, and the type of reconstructed network they provide. For example, SCODE, PPCOR and SINCERITIES infer a directed and signed GRN, which can further be used easily to make dynamical models, if necessary.

**Table 2:** GRN inference tools from single-cell transcriptomics categorized by their inferring algorithm. The characteristics of the inferred network are indicated as follows: ⊘ undirected, ▷ directed, ▶ directed and signed.

| Tool | | Method | Language | Reference |
|---|---|---|---|---|
| SINCERITIES | ▷ | Correlation ensemble | R/Matlab | (Papili Gao et al., 2018) |
| GENIE3 | ▷ | ODE + Regression | R | (Huynh-Thu et al., 2010) |
| GRNBoost2 | ▷ | Regression | Python | (Moerman et al., 2019) |
| PPCOR | ▷ | Semi-partial correlation | R | (Kim, 2015) |
| LEAP | ⊘ | Correlation | R | (Specht and Li, 2016) |
| CARDAMOM* WASABI HARISSA | ▶ | Regression + wave propagation | Python | (Ventre et al., 2022) (Bonnaffoux et al., 2019) (Herbach et al., 2017) |
| SINGE | ▷ | Granger causality | Matlab | (Deshpande et al., 2022) |
| AR1MA1 - VBEM | ▶ | Bayesian Dynamics | Matlab | (Sanchez-Castillo et al., 2018) |
| GRISLI | ▷ | ODE | Matlab | (Aubin-Frankowski and Vert, 2020b) |
| SCODE | ▷ | ODE | R/Julia/Ruby | (Matsumoto et al., 2017) |
| SCRIBE | ▶ | Mutual Information | C++/R | (Qiu et al., 2020) |
| PIDC | ⊘ | Mutual Information | Julia | (Chan et al., 2017) |
| Boolean Pseudotime | ▶ | Boolean Model | Python | (Hamey et al., 2017) |
| InferenceSnapshot | ▶ | Boolean Model | C++/Matlab | (Ocone et al., 2015) |

* Combined methods

## 3. GRN validation

A crucial and essential step in performing GRN inference is validation of the inferred network. Having a validation protocol that allows evaluation of the *score* of each proposed model is very important in order to choose the optimal inference procedure among the variety of existing algorithms. Despite advances in this direction, evaluating the effectiveness of the inference method remains an open challenge, mostly due to limitations in ground truth/gold standard datasets. In some cases, available networks from different public databases, such as RegulonDB (Salgado et al., 2013), KEGG (Kanehisa et al., 2008), ESCAPE (Xu et al., 2013) etc, are used as ground-truth (or reference) networks, and a comparison between the inferred and the reference gold-standard networks is performed.

Particularly, the ESCAPE database (Xu et al., 2013) represents one of the most comprehensive repositories for ChIP-seq - perturbation data. However, the repository contains reference data for a limited number of interactions, limiting the validation to a subset of the inferred network. This implies that the interactions for which the reference data is absent are considered as not existing, raising important questions on the implication of previous biases or the potential to infer novel interactions and regulators. Generally, GRN inference methods aim to infer novel interactions and regulators that might lead to identifying new regulatory pathways involved in the system under consideration, for which gold standard references with high scores are absent or sparse. Another possibility for network validation comes from using simulated data, which can be engineered to include several conditions and measurements (Marbach et al., 2012) - yet the extent of coverage of the inferred interactions remains limited for small network size.

Other than validating the inferred network with a gold-standard one, another important point in GNR inference is network comparison between different inference methods, on the same dataset, thus providing an estimate of the "robustness" of the inferred network. Additionally, one may perform this benchmark analysis in order to choose the algorithm/method that best suits a given dataset.

Quantitatively, the algorithm's performance can be evaluated as for any prediction by two metrics: (1) *area under the precision-recall curve,* estimating the performance of a certain algorithm, and (2) *area under receiver operating characteristic curve,* comparing the GRN inferred by the algorithm against a gold standard network.Considering every inferred interaction in the GRN as a positive (true - TP, or false - FP) and any missing one as negative (true - TN, or false - FN), the two measures are defined as:

1. **Area under precision-recall (AUPR):** starting from the inferred weighted GRN, one might start by filtering by the highest values of the weights ($w_{ij} \in [0, 1]$), for which the *precision* (prediction) will be high and recall (*retrieved*) will be low. Adding interactions with lower weights will lead to a decrease of precision, until at recall = 1 (no filter, fully connected network) the precision will represent the fraction of true positives over the total number of edges in the fully connected network. Graphically, this procedure is represented by a precision-recall curve: a good performing algorithm will have a high precision even with increasing the recall, and an *area under precision-recall* close to 1 (Fig. 4 (a)). Generally, this measure is used as a global performance estimator for most of the inference algorithms.

2. **Area under the receiver operating characteristic (AUROC) curve score**, defined as the ratio between the true positives and false positives, graphically represented with the TP on the $y$ axis and the FP on the $x$ axis (Fig. 4 (b)).

In addition, the structural features of inferred networks, such as PageRank, heat diffusion, the shortest path, etc., can be used to compare GRNs inferred from different methods (Arici and Tuncbag, 2021), whereas other metrics can be used to define the method performance, such as stability (across simulations, artificially removing measurements), identification of network motifs, and computational time and memory usage (Pratapa et al., 2020).

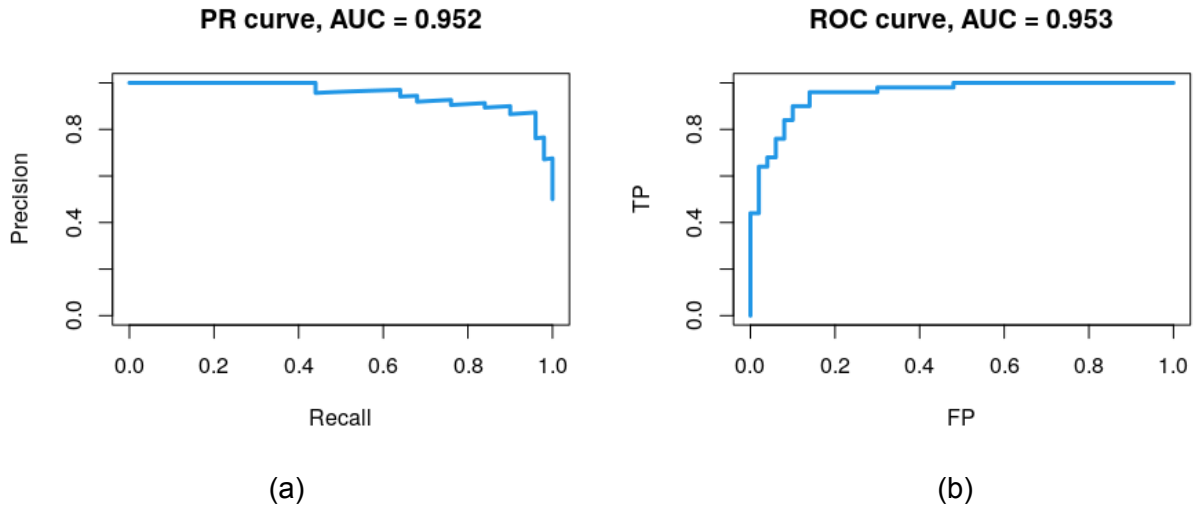

(a) (b)

**Fig. 4:** Example of AUPR (a) and AUROC (b) curves. The graphs are obtained using PRROC R-package (Grau et al., 2015).

An important question arises when comparing the structure of the same cellular network of two different cellular states, e.g different phenotypes of a certain cell or healthy vs disease conditions. This comparative analysis, referred to as *differential network analysis (DiNA)*, remains a challenge, in parallel with the development of GRN inference methods, especially since different data types other than RNAseq can be included in GRN inference in order to increase the accuracy of the inferred network. For example, using single-cell ATAC-seq, patterns of gene expression can be combined with chromatin accessibility profiles, thus identifying cell subpopulations and groups of cells at different developmental stages (Duren et al., 2021). Consequently, several differential network analysis algorithms have been developed, such as DiffRank (Odibat and Reddy, 2011), `dcanr` R package (Bhuva et al., 2019), DiffK (Fuller et al., 2007), DINA (Gambardella et al., 2013), to name a few. We refer the reader to (Shojaie, 2021) for a statistical perspective of DiNA and (Lichtblau et al., 2017) for a DiNA algorithm comparison. In principle, these algorithms consist of combining the information from two main computational tasks: (i) *differential expression analysis*, which estimates the differential gene expression abundance for each gene in the GRN between the two conditions, and (ii) *network expression analysis*, which estimates the importance of each gene in the GRN, based on the topological properties of the network. In this way, DiNA algorithms aim at identifying genes or subnetworks of genes whose expression changes the most across conditions and they are especially suitable in cases when the changes in the network structure lead to phenotypic changes in the system. Therefore, numerous publications have demonstrated the power of this analysis, such as for identifying key TFs involved in cancer, when compared to healthy controls (Duren et al., 2021).

## 4. Applications

Although inferring regulatory networks is a trending topic and despite the multitude of different algorithms available, GRN inference methods struggle to reach a high performance in real-world studies, on both bulk and single-cell RNA-seq data, as reported in (Chen and Mar, 2018). Therefore their application to biologically relevant datasets remains limited. Nevertheless, some applications of inferring GRNs for novel discoveries in biology have been presented in most of the cited works on GRN inference methods. Other applications include building molecular disease maps (Janssens et al., 2022; Van Hove et al., 2019), phenotypic characterization of a cell in a given microenvironment (Lambrechts et al., 2018; Patsalos et al., 2021), identifying predictive or prognostic biomarkers (Lu et al., 2021), discovering new therapeutic targets/regulators (Hossain et al., 2021), performing extensive studies on performance of the available methods on different datasets/conditions (Duren et al., 2021; Jackson et al., 2020; Raharinirina et al., 2021b), and many more.

## 5. From static to dynamic networks

Interestingly, inferred GRNs can serve as a backbone for constructing dynamical models: one can start from a temporal analysis of the system behavior under multiple conditions, describing how the abundances of the genes in the network change due to their regulatory interactions. Dynamical models of GRNs cover a spectrum as wide as the GRN inference methods do. Many dynamical models exist, ranging from continuous quantitative ODE models to discrete logic quantitative models (Fig. 5, (a)). In continuous models, the temporal evolution of the state $X_i$ of gene $i$ ($i = 1, 2,..., N$) in the network is given by a continuous ODE function of its regulators. In this way, the regulatory interactions between genes are modeled as chemical reaction models, or species interactions in ecological models (Polynikis et al., 2009; Takeuchi, 1996). Beneficially, these models can capture the temporal evolution of the system at the level of individual reactions. However, their usage in dynamical modeling of GRN remains limited because of the detailed and complete mathematical and parametric description they require. On the other side of the spectrum, in discrete models, such as multivariate logic models (Aldridge et al., 2009), Petri nets (Murata, 1989), or Boolean models (Glass and Kauffman, 1973; Kauffman and Kauffman, 1993), a discrete logic of interactions is applied and the temporal evolution of expression of each gene is given by a discrete (and often logic) function of their regulators. Contrary to continuous models, discrete models can be applied with no or considerably fewer parameters and fragmented mechanistic description, making them suitable for large networks. However, the system behavior is only described (semi)quantitatively. Other methods use the network structure to infer the asymptotic behavior of the system, i.e, identifying only the steady states of the system, considering only the static GRN and a set of initial conditions, without requiring a dynamical description (Lee and Cho, 2018; Marazzi et al., 2022). For a modeler, the choice of the modeling method is going to depend on the type of questions being asked on the system, the type of available description of the system dynamics (quantitative or qualitative), the type and amount of available data, prior knowledge, etc. Whichever the dynamical model used, the main goal is to identify the phenotypic changes of a cell in response to certain extracellular environmental conditions, drugs, to cellular interactions in the microenvironment, or in experimental knockout/overexpression experiments. Graphically, these phenotypic changes can be identified as the system's steady states or attractors (fixed points or limit cycles) (Fig. 5 (b)) (Shah et al., 2018), defined as a state (a vector of expression values of each gene), which remains unvaried even in the presence of perturbations. Usually, a GRN can have multiple attractors, representing the possible phenotypes or cell states that can be reached when starting from the given initial conditions. In this case, further analysis must be performed on the attractors, to study their stability, their biological relevance, or - additionally - their categorization into known phenotypes.

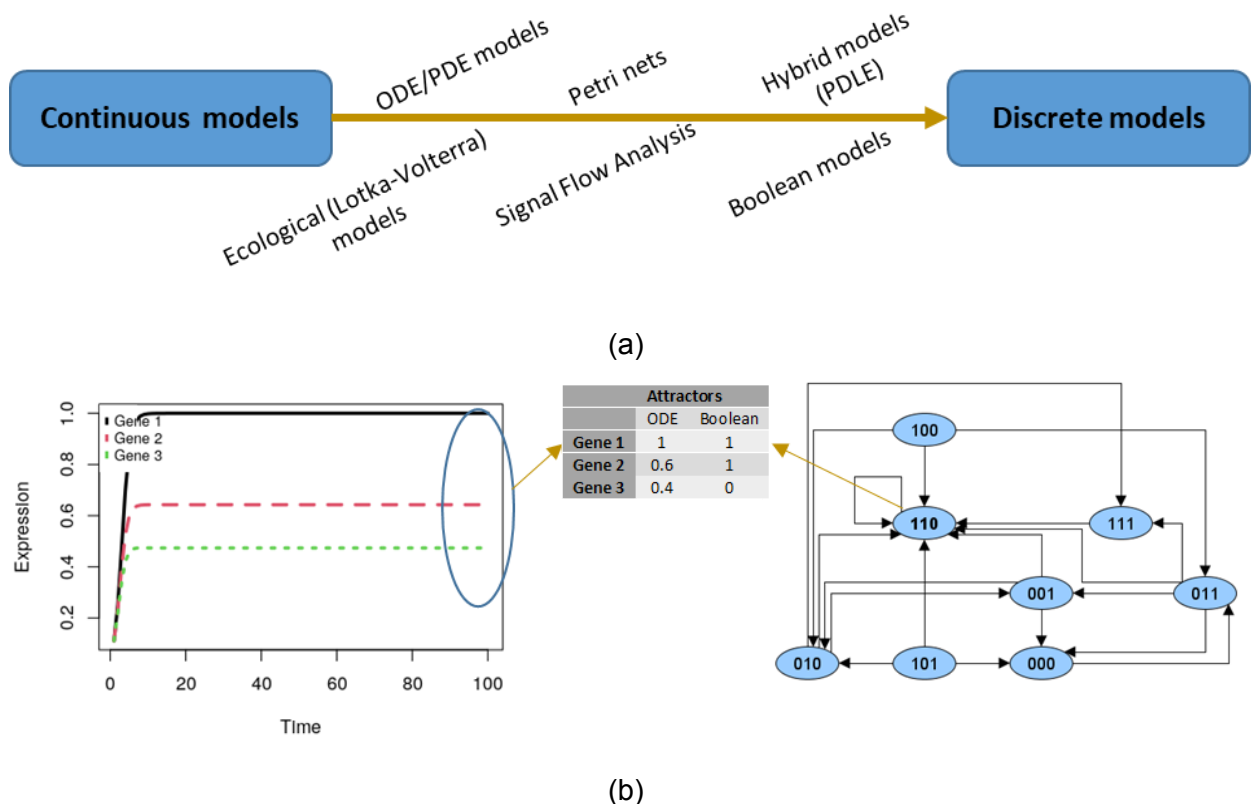

**Fig. 5: (a) Dynamical models of gene regulatory networks**, represented as a spectrum of models ranging from continuous to discrete. **(b) Steady states of a system composed of 3 genes**: (*left*) ODE model, as a system of 3 interacting species, (*right*) Boolean model, as logic-based interacting entities.

Particularly in Boolean models, similar to using time-series of expression data for GRN inference, the temporal information in the changes of the genes' expression is also used to infer the Boolean rules that govern these changes (Barman and Kwon, 2018; Gjerga et al., 2020; Hall and Niarakis, 2021; Henao et al., 2022; Ostrowski et al., 2016; Razzaq et al., 2018), which can be then studied using several available tools (Naldi et al., 2018). More recently, patient-specific Boolean models have been developed to design targeted therapy strategies for patients based on their omics profile (Montagud et al., 2022). Importantly, these intracellular models can be further integrated with other cell population models, like agent-based models or metabolic models, thus providing a multilevel description of the system dynamics, including mechanistic functionalities like cell motility, cytokine diffusion, tissue expansion and spatial organization, etc. (Aguilar et al., 2020; Letort et al., 2019; Stoll et al., 2022). This combined approach enables addressing more complex questions, like drug design, or therapy action from the cell to the tissue scale.

## 6. Discussion - open challenges

In this review we give a basic introduction to GRNs, their topological characteristics and, most importantly, describe the main GRN inference methods. Our aim is to give life scientists an overview of how to use the abstract concept of GRNs to investigate complex cellular molecular interactions more thoroughly and identify how specific interactions determine cells' behavior and response to the environment. With the increasing abundance of transcriptomics data, data-derived GRNs have the potential to capture novel gene interactions and help to expand our knowledge of important molecular pathways. Despite the increasing interest in developing high performing GRN inference methods and this field of research having been active for more than 20 years, often the application of even the best performing methods in real-world studies raises questions about their reliability and purely data-driven GRN inference remains still an open challenge, especially in single-cell RNAseq. Recent promising research addresses the challenge of network inference in the presence of incomplete multi-omics datasets, resulting in the development of advanced computational methods (Henao et al., 2022). We speculate that perhaps the information provided from transcriptomics data, even in single-cell, is not sufficient to cover the complex cellular processes giving rise to phenotypes, and that the inherited concept that co-expression patterns might reveal putative gene interactions might not be universally applicable. For example, due to post-transcription and translation processes, a highly expressed mRNA may not lead to a functional protein, and vice versa - a synthesized protein does not necessarily possess the necessary activation state or conformation or localization to affect its downstream targets. Therefore, there is a necessity to produce multiple types of omics data and the challenge in performing GRN inference will be in how to integrate the data effectively.

Additional limitations in the performance of GRN inference methods can come from other - often ignored - sources. For example, gathering experimental samples from various patients can hinder the variability characterizing patients of individual history, immune system, or genetics. Experimental protocols can present restrictions as well, in limited measurements of expression (transcription, degradation, sequencing capture, etc.), heterogeneity in bulk datasets or incompleteness in single-cell RNAseq transcriptomics.

Another challenge in data-driven GRN inference is their interpretability and the difficulty in dealing with the high complexity increasing with the network size. During the cell state transitions (like differentiation or polarization) in a multicellular system, we can imagine that the interactions between TFs and their target genes can be cell-type specific, thus representing the regulatory network of a specific cell. However, all of the inference methods produce a single

GRN thus failing to provide the dynamics of the underlying mechanisms during cell state transition during the time-series. One possibility in dealing with this challenge is to infer stage-specific GRNs, thus having a time-evolving GRN which can help understanding how the involvement and interactions of specific genes/TFs lead to cell state transition. However, studying and understanding time-evolving GRNs remains an unexplored field of high complexity. Additionally, tracking specific regulatory pathways or interactions rapidly becomes a real challenge when dealing with large networks consisting of thousands of genes. We assume that topological network analysis might help in reducing the network size by selecting the most influencing nodes (by their centrality measures), although there is still a gap in our understanding between the structural and dynamical properties of the network (Luscombe et al., 2004; Klamt et al., 2006; Barabási et al., 2011; Yang et al., 2018).

One of the most difficult challenges in performing data-driven GRN inference is validating the results and estimating the method performance in real-case studies. Benchmarking the inference methods with simulated datasets from a prior knowledge network (PKN) and a small list of genes can be an efficient way to estimate the methods' performance. But this is far from real-case studies, where the actual regulatory mechanisms are mostly unknown and - in some cases - experimentally unexplored. Therefore using public databases (usually built on bulk-RNAseq) for validating the inferred GRNs can be limited to small-size GRNs and narrowed around the highly studied regulatory pathways. Consequently, this leads to biased conclusions and a real difficulty in identifying novel regulatory pathways that might play an important role in the system under study.

We believe that many of these challenges will be addressed in the new inference methods to be developed in the future. Despite the considerable improvements and the rapid growth in numbers of GRN inference methods, this remains a relatively new and highly complex field. Feeding the methods with different types of 'omics data and prior knowledge, GRN inference can help discover unknown pathways, biological components and interactions, thus further increasing our knowledge, in a positive feedback loop.